\documentclass[sigconf]{acmart} 

\usepackage{makecell}
\usepackage{enumitem}
\usepackage{multirow}
\usepackage{subfigure}
\usepackage{graphicx}
\usepackage{subfigure}
\usepackage{caption}
\usepackage{subcaption}
\usepackage{tabularx}
\usepackage{xcolor}

\newif\ifredact
\newif\ifcomment

\redactfalse  
\commenttrue  

\newcommand{\dquote}[1]{\textit{``#1''}}

\ifcomment
  \newcommand{\missing}[1]{\textcolor{red}{~#1}}
  \newcommand{\ken}[1]{\textcolor{magenta}{~Kenny: #1}}
  \newcommand{\wei}[1]{\textcolor{orange}{~Weiyan: #1}}
  \newcommand{\discuss}[1]{\textcolor{purple}{Used in discussion: #1}}
\else
  \newcommand{\missing}[1]{}
  \newcommand{\ken}[1]{}
  \newcommand{\wei}[1]{}
  \newcommand{\discuss}[1]{}
\fi

\AtBeginDocument{%
  }

\begin{document}
\settopmatter{printacmref=false} 
\renewcommand\footnotetextcopyrightpermission[1]{} 
\pagestyle{plain} 

\definecolor{Plum}{rgb}{0.914, 0.008, 0.980}
\definecolor{NavyBlue}{rgb}{0.137, 0.596, 0.969}

\title{Towards Multimodal Large-Language Models for Parent-Child Interaction: A Focus on Joint Attention}

\author{Weiyan Shi}
\email{weiyanshi6@gmail.com}
\orcid{0009-0001-6035-9678}
\affiliation{
  \institution{Singapore University of Technology and Design}
  \country{Singapore}
}

\author{Viet Hai Le}
\email{lehaivin03@gmail.com}
\orcid{0009-0007-3040-6625}
\affiliation{
  \institution{Singapore University of Technology and Design}
  \country{Singapore}
}

\author{Kenny Tsu Wei Choo}
\email{kennytwchoo@gmail.com}
\orcid{0000-0003-3845-9143}
\affiliation{
  \institution{Singapore University of Technology and Design}
  \country{Singapore}
}
\authornote{Corresponding author. This is a preprint of the paper accepted at CHI 2025 Late Breaking Work. The final version will be available in the ACM Digital Library.}

\renewcommand{\shortauthors}{Shi et al.}

\begin{abstract}
Joint attention is a critical component of early speech-language development and a key indicator of effective parent-child interaction.
However, research on detecting and analysing joint attention remains limited, particularly for Multimodal Large Language Models (MLLMs).
This study evaluates MLLMs' ability to comprehend joint attention by analysing 26 parent-child interaction videos annotated by two speech-language pathologists. 
These annotations identify strong and poor joint attention segments, serving as benchmarks for evaluating the models' interpretive capabilities. 
Our findings reveal that current MLLMs struggle to accurately interpret joint attention due to a lack of nuanced understanding of child-initiated eye contact, a crucial component of joint attention dynamics.
This study highlights the importance of incorporating detailed eye contact to enhance MLLMs' multimodal reasoning.
Addressing these gaps is essential for future research to advance the use of MLLMs in analysing and supporting parent-child interactions.
\end{abstract}



\maketitle

\section{INTRODUCTION}
In parent-child interaction, joint attention, play, and imitation are crucial for promoting child speech-language development~\cite{charman2000testing}. Joint attention, in particular, plays a foundational role in fostering communication and language skills. Two individuals simultaneously focus on the same object or event, sharing their experience through eye contact, gestures, or verbal expressions~\cite{hanen_joint_attention}. Such mechanisms facilitate shared focus and enable parents and infants to achieve the social coordination necessary for language learning~\cite{baldwin2014understanding}.

While some technologies~\cite{hwang2014talkbetter,song2016talklime} have been developed to enhance joint attention, there is little work~\cite{kwon2022captivate} on detecting joint attention in parent-child interaction. With the rise of multimodal large language models (MLLMs), which have demonstrated strengths in processing conversations and analysing multimodal content~\cite{lu2024gpt}, these models offer promising tools to identify joint attention segments in parent-child videos.

To bridge this gap, our study explores the use of MLLMs to understand joint attention in parent-child interactions. Using our selection criteria, we collected a final dataset of 26 publicly sourced online videos showcasing parent-child interactions. Collaborating with two professional speech-language pathologists (SLPs), we annotated the videos with strong and poor joint attention segments. By comparing the MLLMs' outputs with expert annotations, we identified the limitations of joint attention detection in current MLLMs. 

Our contributions are as follows:

\begin{itemize}
    \item We collected and labelled the first-of-its-kind dataset of parent-child interaction videos. This was labelled for joint attention together with professional SLPs.
    \item We tested the capabilities of MLLMs in understanding joint attention within these interactions and identified the limitations of current MLLMs' understanding of joint attention for speech-language therapy.
\end{itemize}
\section{RELATED WORK}

\subsection{Parent-Child Interaction Challenges in Early Speech-Language Development}

Parent-child interactions, especially joint attention activities, are essential for speech-language development. When parents lack proper guidance or inadequately follow the advice of SLPs, children's speech-language development can be affected--leading to poorer outcomes for children who already have communication developmental delays. High costs and a shortage of SLPs, particularly in rural or underserved areas, often limit access to professional speech therapy~\cite{o2005barriers}. Programs like DIR Floortime~\cite{dir_floortime}, It Takes Two to Talk~\cite{pepper2004talk}, and More Than Words~\cite{sussman1999words} have been developed to help parents promote joint attention and language development. DIR Floortime emphasises emotional connections and following the child's lead~\cite{dir_floortime}. The Hanen Centre's It Takes Two to Talk~\cite{pepper2004talk} and More Than Words~\cite{sussman1999words} programs provide practical strategies to improve communication. Building on these efforts, integrating technology to detect and analyse joint attention could further enhance parent-child interactions and support speech-language development.

\subsection{Supporting Parent-Child Interaction}
Recent technological advancements have introduced systems aimed at enhancing parent-child interactions.
Chan et al. developed WAKEY, a system designed to improve parent-child communication during morning routines, enabling parents and preschool children to start their day more smoothly~\cite{chan2017wakey}. However, WAKEY relies entirely on manual input and logging by parents for tasks such as scheduling and tracking phrase usage frequency, lacking the ability to automatically gather contextual data, such as audio or video interaction details.
Hwang et al. proposed TalkBetter, a system that analyses turn-taking and provides real-time feedback and tailored recommendations to help parents foster their children's language development~\cite{hwang2014talkbetter}. Song et al. introduced TalkLIME, a mobile system that enhances parent-child interactions by providing real-time feedback on metrics like utterance count, turn-taking, and initiation ratio~\cite{song2016talklime}. 
However, both TalkBetter and TalkLIME rely solely on voice-based information, lacking conversational or visual context, which restricts their ability to capture the nuances of parent-child interactions fully.
Jeong et al. developed Captivate!, a system that uses multimodal sensing, including gaze estimation and speech recognition, to detect joint attention and recommend phrases during parent-child interactions~\cite{kwon2022captivate}. While it can help determine joint attention on a specific object, it cannot provide a detailed contextual understanding of interactions beyond this.
Existing technologies in this domain often lack either multimodal information or the ability to fully understand the context of interactions, highlighting the need for systems that integrate richer multimodal data and contextual awareness to support parent-child interactions better.


\subsection{Potential of MLLMs in Parent-Child Interaction}

Large Language Models (LLMs) have been increasingly applied to enhance human-human interactions, showcasing their potential to facilitate seamless, context-aware communication. 
Tanneberg et al.~\cite{tanneberg2024attentivesupport} proposed a robot-based system that leverages LLMs for unobtrusive group support. 
By analysing spoken dialogue and contextual cues, the system determines when to intervene and provides targeted assistance, such as correcting wrong information or supporting task completion. This work highlights the potential of LLMs to foster collaborative efficiency in group settings.
Nihei et al.~\cite{nihei2024chatbots} developed CMBot, an LLM-powered system to mediate communication between older adults and their families.
The system dynamically adjusts self-disclosure levels and generates personalised conversation prompts based on themes like hobbies or finances. 
By referencing prior interactions, CMBot avoids repetition and provides contextually relevant recommendations, enhancing intergenerational communication.
Building on these advancements, MLLMs further extend the capabilities of traditional LLMs by integrating text, visual, and auditory modalities. This multimodal integration enables a deeper understanding of complex interactions across diverse environments.
For example, models like GPT-4o excel in tasks like video temporal grounding, dense captioning, and summarization~\cite{lu2024gpt}.
Given these advancements, applying MLLMs to detect and analyse joint attention in parent-child interactions represents a promising frontier.
By leveraging their ability to process multimodal inputs, MLLMs could offer innovative solutions for capturing and understanding the nuances of shared attention dynamics and more in real-world contexts.
\begin{table*}[htb!]
\centering
\caption{Comparison of annotation between SLPs: \textbf{SLP} (annotator ID), \textbf{Total} (number of segments), \textbf{Strong (\%)} and \textbf{Poor (\%)} (counts and percentages of strong/poor joint attention), \textbf{Avg. (s)} (average segment duration), \textbf{Avg. Strong (s)}, and \textbf{Avg. Poor (s)} (average durations of strong and poor segments). The \textbf{Intersection} row shows metrics for mutually agreed segments.}
\begin{tabular}{p{0.12\textwidth} p{0.12\textwidth} p{0.12\textwidth} p{0.12\textwidth} p{0.12\textwidth} p{0.12\textwidth} p{0.12\textwidth}} 
\toprule
\textbf{SLP} & \textbf{Total} & \textbf{Strong (\%)} & \textbf{Poor (\%)} & \textbf{Avg. (s)} & \textbf{Avg. Strong (s)} & \textbf{Avg. Poor (s)} \\ 
\midrule
S1           & 96             & 89 (92.71)           & 7 (7.29)           & 5.61               & 5.24                      & 10.30                   \\
S2           & 72             & 56 (77.78)           & 16 (22.22)         & 14.51              & 14.85                     & 13.33                   \\
\midrule
Intersection & 62             & 58 (93.55)           & 4 (6.45)           & 3.88               & 3.65                      & 10.73                   \\
\bottomrule
\end{tabular}
\label{tab:segment_metrics}
\end{table*}
\section{Video Data Collection and Preparation}
\subsection{Video Selection Method}
We collected videos from YouTube using targeted keywords \dquote{parent-child interaction} to create a dataset for analysing parent-child interactions, carefully selecting videos that met quality and relevance criteria. Each video had to feature one child and one adult as the primary subjects. While most videos required active interactions between the adult and the child, we also included cases where the adult's role was limited to accompanying a very young child for support without directly engaging in the interaction. Videos involving multiple children or adults actively participating were excluded to maintain the focus on dyadic dynamics.

We applied stringent selection criteria to ensure the videos were suitably focused towards analysing multimodal signals. The scenes needed to be static with minimal camera movement, and the videos had to provide clear views of both the child's and adult's faces, enabling precise analysis of gaze direction and expressions. High-quality audio and visual clarity were also essential for accurate verbal and non-verbal communication observations.

Additionally, the dataset was curated to include children across a broad age range to capture a variety of developmental stages. We prioritised videos that showcased diverse and meaningful interactions, such as language learning activities, skill-building tasks, or natural daily exchanges. All textual elements, including subtitles and transitions, were removed to preserve the raw multimodal content.

\subsection{Dataset Composition and Analysis}
We curated a final dataset comprising 26 parent-child interaction videos that satisfied our quality and relevance criteria. These videos are of dialogue and interaction in fixed settings, covering age ranges between 0-8 years (age information obtained from the descriptions provided on the video websites). The majority of videos focus on children aged 4–6 years (n=16), with fewer videos in the 2–4 (n=5) and 0–2 (n=4) age groups. Only one video includes children aged 6–8, concentrating on younger age groups where parent-child interactions are more actively studied.

The duration of the videos ranges from brief clips of around 0.5 to 12 minutes. Most videos are relatively short: 13 videos are less than 1 minute, and 11 fall between 1 and 5 minutes. Only two videos extend beyond 5 minutes, one categorised as 5–10 minutes and the other over 10 minutes. This distribution highlights the concise nature of the interactions and tasks typically observed in parent-child settings.

The video content in the dataset can be categorised into three main categories. The first category is behaviour guidance and skill modelling (n=10), where parents demonstrate specific techniques (e.g., PRIDE skills, \dquote{Big Ignore} techniques) to guide children's behaviour and improve interaction quality. The second category focuses on language and cognitive development (n=10), showcasing how parent-child interactions foster language learning and cognitive skills through tasks or experiments, such as discussing specific topics or engaging in Piaget's cognitive experiments. The third category involves daily life skills and interaction (n=6), presenting natural exchanges in structured settings, such as reviewing reward charts or ending special playtime, emphasising practical life skills and building strong parent-child relationships.


\section{Speech-Language Pathologists Annotation Study}
\subsection{Annotation Process}
To evaluate the performance of MLLMs in identifying strong or poor joint attention segments, we recruited two SLPs (both female) with extensive experience to provide expert annotations. They each have 7–9 years of experience and training in programs such as DIR Floortime~\cite{dir_floortime}, It Takes Two to Talk~\cite{pepper2004talk}, and More Than Words~\cite{sussman1999words}. While a third SLP was recruited, the session was incomplete and the results were not included in this paper. Our studies complied with ethical regulations approved by our institution's ethics review board.



The study was conducted in person over one month. Each session lasted approximately 2 hours and was held in a private meeting room to ensure focus and comfort. The labelling was carried out using our custom-built annotation system. Videos can be played, paused, or navigated frame-by-frame, while the interactive timeline enables users to select segments and label them as \dquote{strong} or \dquote{poor}. The system exports the results in CSV format, including each labelled segment's start and end times, along with corresponding indices. Additionally, a panel allows experts to add notes during the labelling process.


\subsection{Annotation Outcomes Analysis}
We conducted full annotation interviews with SLP 1 (\textit{S1}) and SLP 2 (\textit{S2}). Through these interviews, a shared criterion for determining strong joint attention was established: it is characterized by the child maintaining consistent eye contact and \textit{active engagement} with the parent during the interaction. 

Table \ref{tab:segment_metrics} summarizes the annotation metrics provided by the two SLPs. \textit{S1} annotated a total of 96 segments. Among these, \textit{92.71\%} were identified as strong segments, with an average segment duration of \textit{5.61 s}. Strong segments were notably shorter, averaging \textit{5.24 s}, while poor segments were longer at \textit{10.30 s}. \textit{S2} annotated 72 segments. Among these, \textit{77.78\%} were classified as strong, and \textit{22.22\%} as poor. The average duration for strong and poor segments was \textit{14.85 s} and \textit{13.33 s}, respectively, reflecting S2's longer segments in annotations.

To ensure fairness and consistency, we calculated the intersection of annotations where SLPs agreed. This resulted in \textit{62} segments, with \textit{93.55\%} identified as strong and \textit{6.45\%} as poor. The average duration for strong and poor segments in the intersection was \textit{3.65 s} and \textit{10.73 s}, respectively, while the overall average duration was \textit{3.88 s}.

\section{MLLM Comparative Study}
\subsection{Experimental Setup and Evaluation Metrics}

To test the performance of MLLMs on our parent-child joint-attention detection task, a fine-grained temporal video grounding challenge, we selected three state-of-the-art MLLMs to evaluate: GPT-4o-2024-08-06\footnote{\url{https://openai.com/index/hello-gpt-4o/}}, Gemini 1.5 Flash\footnote{\url{https://deepmind.google/technologies/gemini/flash/}}, and Video-ChatGPT~\cite{maaz2023video}.

The Gemini 1.5 model directly supports audio and video processing, allowing us to input the entire video seamlessly. In contrast, Video-ChatGPT can only process videos without audio, so we provided the full video without additional preprocessing. GPT-4o, however, does not natively support direct video processing. To address this, we extracted video frames and converted them into a Base64-encoded array of images. 
We also used WhisperX~\cite{bain2023whisperx} to transcribe the videos' audio and manually enriched these transcriptions with notes for greater informativeness. 
For instance, we included annotations such as \dquote{the child is laughing} and descriptions of background sounds, which helped compensate for the unclear speech often observed in younger children. These transcriptions were used as text input, serving as a substitute for audio input for both Video-ChatGPT and GPT-4o.

We designed Instruction Templates (see Table~\ref{tab:instructions}) inspired by Liu et al.~\cite{liu2024bench} and based on our interviews with SLPs. 
Our interviews identified key criteria for determining joint attention, such as consistent eye contact and full engagement between the child and parent.

\begin{table*}[!htb]
\small
  \caption{Instruction templates in our task. \textcolor{orange}{<time>} denotes the timestamp representation, e.g., "23.6s" \textcolor{NavyBlue}{<description>} denotes the brief summary of the segment focused on the child's joint attention behaviour. \textcolor{Plum}{<label>} denotes the child joint attention quality of the segment.}
  \label{tab:instructions}
  \Description{}
  \begin{tabular}{p{.5\linewidth} p{.4\linewidth}}
    \toprule
    \textbf{Instruction} &  \textbf{Example Response}\\
    \midrule
    You are given a video about parent-child interactions. Watch the video carefully and your task is to:

    1. Identify the key moments where the child's joint attention occurs. Child joint attention is defined as the child making eye contact and being fully engaged with the parent through interaction.
    
    2. Specify the timestamps for when each moment starts and ends.

    3. Classify the quality of the child's joint attention into "Strong" and "Poor".

    For example:
    
    - Timestamp: 23.6s - 26.8s
    
    - Description: The child looks at his mother as she gives him instructions.
    
    - Quality: Strong
    &
    Timestamp: \textcolor{orange}{<time>} - \textcolor{orange}{<time>} and \textcolor{orange}{<time>} - \textcolor{orange}{<time>}
    
    Description: \textcolor{NavyBlue}{<description>}
    
    Quality: \textcolor{Plum}{<label>}
    \\
  \bottomrule
\end{tabular}
\end{table*}

\subsubsection{Evaluation Method of MLLMs Output Quality and Content Objectivity}

To assess the objectivity and quality of the models' responses, we define four metrics building on key aspects emphasized by SLPs in their annotations:

\begin{itemize}
    \item \textbf{Time Sensitivity}: Proportion of responses that include \textcolor{orange}{<time>} - \textcolor{orange}{<time>}.
    \item \textbf{Description Accuracy}: Proportion of accurate \textcolor{NavyBlue}{<description>}s within the specified \textcolor{orange}{<time>} - \textcolor{orange}{<time>} in the responses.
    \item \textbf{Eye-Contact Sensitivity}: Proportion of accurate \textcolor{NavyBlue}{<description>} that include eye contact information.
    \item \textbf{Eye-Contact Accuracy}: Proportion of accurate \textcolor{NavyBlue}{<description>} that include accurate eye contact information.
\end{itemize}

We reviewed the responses against the original videos. For each response,  we verified the inclusion of time information, descriptions' accuracy, and eye-contact events' presence and correctness. Annotations were assigned based on these criteria, ensuring a reliable benchmark for comparing the models' performance across the four evaluation metrics.

\subsubsection{Evaluation Method of MLLMs Temporal Grounding in Joint Attention Segments}

To evaluate the temporal understanding capabilities of MLLMs in detecting strong and poor joint attention segments, we used mean Intersection over Union (mIoU) and Recall at IoU thresholds (R@m). These metrics assess the alignment quality between predicted and ground truth time segments. IoU measures the overlap between the predicted and ground truth time segments, defined as the ratio of the intersection to the union of these two time intervals. A higher IoU indicates better alignment, and mIoU represents the average IoU across all segments. R@m evaluates the proportion of ground truth segments with at least one predicted segment with sufficient overlap based on predefined IoU thresholds.

The evaluation process used the intersection of annotations from the two SLPs as the ground truth. Each model outputted its predicted timestamps for strong and poor joint attention segments. Using these predictions, we computed the metrics as follows:
\begin{itemize}
    \item For \textbf{mIoU}, we calculated the average overlap between the predicted and ground truth timestamps across all segments, ensuring that both temporal precision and alignment quality were considered.
    \item For \textbf{R@m}, we assessed how many ground truth segments had at least one predicted segment with sufficient overlap (defined by IoU thresholds of 0.3, 0.5, and 0.7).
\end{itemize}

\subsection{Performance Analysis and Findings}
We conducted experiments on three models. For GPT-4o and Gemini-1.5-Flash, we utilized their respective APIs. The model was deployed and executed locally on a Ubuntu 22.04 LTS server equipped with two NVIDIA RTX 3090 GPUs for Video-ChatGPT. 

\subsubsection{Evaluation Result of MLLMs response Quality and Content Objectivity}

Table~\ref{tab:model-evaluation} presents the evaluation of three models—GPT-4o, Gemini-1.5-Flash, and Video-ChatGPT—across four metrics.
For \textit{Time Sensitivity}, both GPT-4o and Gemini-1.5-Flash achieved perfect scores (\textit{100\%}), indicating their consistent inclusion of time information in the responses. 
In contrast, Video-ChatGPT performed poorly, with only a small proportion of responses with time information (\textit{36.36\%}).
GPT-4o demonstrated the strongest performance (\textit{90.17\%}) in \textit{Description Accuracy}, accurately describing tasks and scenes within the specified time ranges. 
Gemini-1.5-Flash and Video-ChatGPT performed worse, scoring (\textit{50.53\%}) and (\textit{35.71\%}), respectively. 
Gemini-1.5-Flash frequently misinterpreted interactions, often assuming that the child and parent had eye contact. At the same time, Video-ChatGPT exhibited significant scene misrecognition, such as mistaking a boy for both a boy and a girl.
For \textit{Eye-Contact Sensitivity}, GPT-4o showed limited inclusion of eye-contact information (\textit{23.64\%}), while Gemini-1.5-Flash demonstrated a much higher sensitivity (\textit{92.12\%}), actively identifying eye-contact cues. Video-ChatGPT failed to provide any eye-contact information in its responses.
For \textit{Eye-Contact Accuracy}, GPT-4o achieves the best performance (\textit{62.82\%}), avoiding overestimations of eye contact. Gemini-1.5-Flash has a lower accuracy (\textit{43.92\%}) due to frequent overestimation, assuming eye contact occurs more often than it does. Video-ChatGPT does not output eye-contact information, making it unsuitable for tasks requiring such data.

\begin{table*}[htb!]
\centering
\caption{Evaluation of MLLMs Output Quality and Content Objectivity Across Four Metrics: Time Sensitivity (proportion of responses including time information), Description Accuracy (accuracy of descriptions within specified time ranges), Eye-Contact Sensitivity (proportion of responses including eye contact information), and Eye-Contact Accuracy (accuracy of identified eye-contact information).}
\label{tab:model-evaluation}
\begin{tabular}{p{0.15\textwidth} p{0.18\textwidth} p{0.18\textwidth} p{0.18\textwidth} p{0.18\textwidth}}     
    \toprule
    \textbf{Model} & \textbf{Time Sens.} & \textbf{Description Acc.} & \textbf{Eye-Contact Sens.} & \textbf{Eye-Contact Acc.} \\
    \midrule
    GPT-4o & 100\% & 90.17\% & 23.64\% & 62.82\% \\
    Gemini-1.5-Flash & 100\% & 50.53\% & 92.12\% & 43.92\% \\
    Video-ChatGPT & 36.36\% & 35.71\% & 0\% & 0\% \\
    \bottomrule
\end{tabular}
\end{table*}

While GPT-4o and Gemini-1.5-Flash perform well in Time Sensitivity and Description Accuracy, all models show weaknesses in Eye-Contact Sensitivity and Accuracy. Since Video-ChatGPT performed poorly in time sensitivity, only GPT-4o and Gemini-1.5-Flash were considered for the temporal grounding evaluation in the following section.

\begin{table*}[htb!]
\centering
\caption{Comparison of Temporal Grounding Metrics between Gemini-1.5-Flash and GPT-4o for Strong, Poor, and Overall Joint Attention Segments. The metrics include Recall at IoU thresholds (R@0.3, R@0.5, and R@0.7) and mean Intersection over Union (mIoU). R@0.3, R@0.5, and R@0.7 represent the proportion of ground truth segments with at least one predicted segment with sufficient overlap at IoU thresholds of 0.3, 0.5, and 0.7, respectively. mIoU measures the average alignment quality between predicted and ground truth segments. "Strong" refers to the model's performance on segments with strong joint attention, "Poor" refers to segments with poor joint attention, and "Overall" reflects performance across all segments.}
\begin{tabular}{p{0.25\textwidth} p{0.15\textwidth} p{0.15\textwidth} p{0.15\textwidth} p{0.15\textwidth}}     \toprule
\textbf{Model}               & \textbf{R@0.3 (\%)} & \textbf{R@0.5 (\%)} & \textbf{R@0.7 (\%)} & \textbf{mIoU (\%)} \\ \midrule
GPT-4o (Strong)              & 2.01                 & 0.57                 & 0.57                 & 2.19               \\
Gemini-1.5-Flash (Strong)    & 1.72                 & 0.69                 & 0.52                 & 1.39               \\
GPT-4o (Poor)                & 0.00                 & 0.00                 & 0.00                 & 3.30               \\
Gemini-1.5-Flash (Poor)      & 0.00                 & 0.00                 & 0.00                 & 2.28               \\
\midrule
GPT-4o (Overall)             & 1.77                 & 0.51                 & 0.51                 & 2.29               \\
Gemini-1.5-Flash (Overall)   & 1.17                 & 0.47                 & 0.35                 & 1.78               \\
\bottomrule
\end{tabular}
\label{tab:metrics_comparison}
\end{table*}

\subsubsection{Evaluation Result of MLLMs Temporal Grounding in Joint Attention Segments}

Table \ref{tab:metrics_comparison} highlights that GPT-4o outperforms Gemini-1.5-Flash on strong joint attention segments, achieving a higher mIoU (2.19\% vs. 1.39\%) and a slightly better R@0.3 (2.01\% vs. 1.72\%). This suggests GPT-4o aligns more effectively with ground truth for strong segments, likely due to its broader contextual understanding and reduced reliance on explicit eye-contact information.

For poor joint attention segments, GPT-4o also achieves a higher mIoU (3.30\% vs. 2.28\%), despite both models reporting zero Recall across all IoU thresholds (R@0.3, R@0.5, R@0.7). This discrepancy occurs because mIoU captures partial overlaps between predicted and ground truth regions, even if the overlaps are insufficient to meet recall thresholds. In the poor category, models may predict larger or misaligned regions, contributing to mIoU but failing to exceed the recall thresholds.

Overall, GPT-4o demonstrates superior performance, with a higher mIoU (2.29\% vs. 1.78\%) and slightly better recall values across all categories. However, this low value indicates limited alignment between predicted and ground truth regions. The result reflects the challenges of temporal grounding in joint attention tasks, where precise boundary prediction is difficult.

Despite these results, the findings highlight limitations in current MLLMs for direct understanding of joint attention. As noted earlier, deficiencies in processing eye-contact information significantly reduce accuracy, underscoring the need for improved multimodal integration.
\section{DISCUSSION AND FUTURE WORK}



This work focuses on joint attention in parent-child interactions, exploring the potential and limitations of MLLMs in detecting and analysing this critical concept. Additionally, our findings provide insights that may inform future research on other important aspects of parent-child interactions, such as imitation and joint intention.

\subsection{Towards better Human-Centered AI for Joint Attention}
Our results highlight significant gaps in MLLMs' sensitivity and accuracy regarding eye contact information, a critical element in understanding joint attention. For example, determining what a child is looking at often requires precise gaze estimation and object detection, areas where traditional computer vision methods excel~\cite{ke2024tail, ke2025detection, cyu2025identifying}. A potential direction for researchers is to explore how eye contact information can be effectively integrated into MLLMs by combining these conventional approaches with multimodal capabilities. Another challenge is improving MLLMs' performance on complex tasks like temporal grounding for joint attention, which requires a deeper contextual and sequential understanding of interactions.

\subsection{Need for Dataset Diversification}
Our findings emphasize the critical need to diversify datasets to advance parent-child interaction research. Currently, available datasets are primarily focused on specific, fixed scenarios, limiting their generalizability. To address this, researchers should prioritize collecting data from a broader range of parent-child interaction contexts. For example, beyond the fixed scenarios we collected, datasets capturing interactions during free activities within a room or other everyday environments could provide richer insights. This would not only support research on joint attention but also facilitate analysis of other key components, such as joint intention and imitation. Moreover, addressing the scarcity of "poor" joint attention segments in existing datasets is crucial for enabling more robust model training and evaluation. Expanding the variety and scope of datasets will be instrumental in enhancing the effectiveness and applicability of computational models in this domain.

By addressing these challenges and advancing both dataset development and model capabilities, future research can better align MLLMs with expert-level reasoning in parent-child interaction analysis.
\section{CONCLUSION}
Joint attention plays a vital role in early language development, serving as a cornerstone for effective parent-child interaction. In this study, we analysed 26 parent-child interaction videos, with annotations from two SLPs identifying segments of strong and poor joint attention. 

We tested the capabilities of MLLMs in understanding these interactions. Our findings reveal that current MLLMs face limitations due to the absence of explicit eye-contact information, which significantly impacts their ability to comprehend joint attention.

\begin{acks}
This research is supported by the National Research Foundation, Singapore, under its AI Singapore Programme (AISG Award No. AISG2-TC-2022-007).
\end{acks}

\bibliographystyle{ACM-Reference-Format}
\bibliography{main}


\begin{thebibliography}{20}


\ifx \showCODEN    \undefined \def \showCODEN     #1{\unskip}     \fi
\ifx \showDOI      \undefined \def \showDOI       #1{#1}\fi
\ifx \showISBNx    \undefined \def \showISBNx     #1{\unskip}     \fi
\ifx \showISBNxiii \undefined \def \showISBNxiii  #1{\unskip}     \fi
\ifx \showISSN     \undefined \def \showISSN      #1{\unskip}     \fi
\ifx \showLCCN     \undefined \def \showLCCN      #1{\unskip}     \fi
\ifx \shownote     \undefined \def \shownote      #1{#1}          \fi
\ifx \showarticletitle \undefined \def \showarticletitle #1{#1}   \fi
\ifx \showURL      \undefined \def \showURL       {\relax}        \fi
\providecommand\bibfield[2]{#2}
\providecommand\bibinfo[2]{#2}
\providecommand\natexlab[1]{#1}
\providecommand\showeprint[2][]{arXiv:#2}

\bibitem[Bain et~al\mbox{.}(2023)]%
        {bain2023whisperx}
\bibfield{author}{\bibinfo{person}{Max Bain}, \bibinfo{person}{Jaesung Huh}, \bibinfo{person}{Tengda Han}, {and} \bibinfo{person}{Andrew Zisserman}.} \bibinfo{year}{2023}\natexlab{}.
\newblock \showarticletitle{Whisperx: Time-accurate speech transcription of long-form audio}. In \bibinfo{booktitle}{\emph{Proceedings of the International Speech Communication Association (INTERSPEECH)}}. \bibinfo{pages}{4489--4493}.
\newblock


\bibitem[Baldwin(2014)]%
        {baldwin2014understanding}
\bibfield{author}{\bibinfo{person}{Dare~A Baldwin}.} \bibinfo{year}{2014}\natexlab{}.
\newblock \showarticletitle{Understanding the link between joint attention and language}.
\newblock In \bibinfo{booktitle}{\emph{Joint attention}}. \bibinfo{publisher}{Psychology Press}, \bibinfo{pages}{131--158}.
\newblock


\bibitem[Chan et~al\mbox{.}(2017)]%
        {chan2017wakey}
\bibfield{author}{\bibinfo{person}{Meng-Ying Chan}, \bibinfo{person}{Yi-Hsuan Lin}, \bibinfo{person}{Long-Fei Lin}, \bibinfo{person}{Ting-Wei Lin}, \bibinfo{person}{Wei-Che Hsu}, \bibinfo{person}{Chia-yu Chang}, \bibinfo{person}{Rui Liu}, \bibinfo{person}{Ko-Yu Chang}, \bibinfo{person}{Min-hua Lin}, {and} \bibinfo{person}{Jane Yung-jen Hsu}.} \bibinfo{year}{2017}\natexlab{}.
\newblock \showarticletitle{WAKEY: assisting parent-child communication for better morning routines}. In \bibinfo{booktitle}{\emph{Proceedings of the 2017 ACM Conference on Computer Supported Cooperative Work and Social Computing}}. \bibinfo{pages}{2287--2299}.
\newblock


\bibitem[Charman et~al\mbox{.}(2000)]%
        {charman2000testing}
\bibfield{author}{\bibinfo{person}{Tony Charman}, \bibinfo{person}{Simon Baron-Cohen}, \bibinfo{person}{John Swettenham}, \bibinfo{person}{Gillian Baird}, \bibinfo{person}{Antony Cox}, {and} \bibinfo{person}{Auriol Drew}.} \bibinfo{year}{2000}\natexlab{}.
\newblock \showarticletitle{Testing joint attention, imitation, and play as infancy precursors to language and theory of mind}.
\newblock \bibinfo{journal}{\emph{Cognitive development}} \bibinfo{volume}{15}, \bibinfo{number}{4} (\bibinfo{year}{2000}), \bibinfo{pages}{481--498}.
\newblock


\bibitem[Hwang et~al\mbox{.}(2014)]%
        {hwang2014talkbetter}
\bibfield{author}{\bibinfo{person}{Inseok Hwang}, \bibinfo{person}{Chungkuk Yoo}, \bibinfo{person}{Chanyou Hwang}, \bibinfo{person}{Dongsun Yim}, \bibinfo{person}{Youngki Lee}, \bibinfo{person}{Chulhong Min}, \bibinfo{person}{John Kim}, {and} \bibinfo{person}{Junehwa Song}.} \bibinfo{year}{2014}\natexlab{}.
\newblock \showarticletitle{TalkBetter: family-driven mobile intervention care for children with language delay}. In \bibinfo{booktitle}{\emph{Proceedings of the 17th ACM conference on Computer supported cooperative work \& social computing}}. \bibinfo{pages}{1283--1296}.
\newblock


\bibitem[Ke and Yin(2024)]%
        {ke2024tail}
\bibfield{author}{\bibinfo{person}{Zong Ke} {and} \bibinfo{person}{Yuchen Yin}.} \bibinfo{year}{2024}\natexlab{}.
\newblock \showarticletitle{Tail Risk Alert Based on Conditional Autoregressive VaR by Regression Quantiles and Machine Learning Algorithms}. In \bibinfo{booktitle}{\emph{2024 5th International Conference on Artificial Intelligence and Computer Engineering (ICAICE)}}. IEEE, \bibinfo{pages}{527--532}.
\newblock


\bibitem[Ke et~al\mbox{.}(2025)]%
        {ke2025detection}
\bibfield{author}{\bibinfo{person}{Zong Ke}, \bibinfo{person}{Shicheng Zhou}, \bibinfo{person}{Yining Zhou}, \bibinfo{person}{Chia~Hong Chang}, {and} \bibinfo{person}{Rong Zhang}.} \bibinfo{year}{2025}\natexlab{}.
\newblock \showarticletitle{Detection of AI Deepfake and Fraud in Online Payments Using GAN-Based Models}.
\newblock \bibinfo{journal}{\emph{arXiv preprint arXiv:2501.07033}} (\bibinfo{year}{2025}).
\newblock


\bibitem[Kwon et~al\mbox{.}(2022)]%
        {kwon2022captivate}
\bibfield{author}{\bibinfo{person}{Taeahn Kwon}, \bibinfo{person}{Minkyung Jeong}, \bibinfo{person}{Eon-Suk Ko}, {and} \bibinfo{person}{Youngki Lee}.} \bibinfo{year}{2022}\natexlab{}.
\newblock \showarticletitle{Captivate! contextual language guidance for parent--child interaction}. In \bibinfo{booktitle}{\emph{Proceedings of the 2022 CHI Conference on Human Factors in Computing Systems}}. \bibinfo{pages}{1--17}.
\newblock


\bibitem[Liu et~al\mbox{.}(2024)]%
        {liu2024bench}
\bibfield{author}{\bibinfo{person}{Ye Liu}, \bibinfo{person}{Zongyang Ma}, \bibinfo{person}{Zhongang Qi}, \bibinfo{person}{Yang Wu}, \bibinfo{person}{Ying Shan}, {and} \bibinfo{person}{Chang~Wen Chen}.} \bibinfo{year}{2024}\natexlab{}.
\newblock \showarticletitle{Et Bench: Towards Open-Ended Event-Level Video-Language Understanding}. In \bibinfo{booktitle}{\emph{Proceedings of the NeurIPS Datasets and Benchmarks Track 2024}}.
\newblock


\bibitem[Lowry(2022)]%
        {hanen_joint_attention}
\bibfield{author}{\bibinfo{person}{Lauren Lowry}.} \bibinfo{year}{2022}\natexlab{}.
\newblock \bibinfo{title}{Paying Attention to Children’s Joint Attention}.
\newblock \bibinfo{howpublished}{\url{https://www.hanen.org/information-tips/paying-attention-to-childrens-joint-attention}}.
\newblock
\newblock
\shownote{Accessed: 2025-01-19}.


\bibitem[Lu et~al\mbox{.}(2024)]%
        {lu2024gpt}
\bibfield{author}{\bibinfo{person}{Chaochao Lu}, \bibinfo{person}{Chen Qian}, \bibinfo{person}{Guodong Zheng}, \bibinfo{person}{Hongxing Fan}, \bibinfo{person}{Hongzhi Gao}, \bibinfo{person}{Jie Zhang}, \bibinfo{person}{Jing Shao}, \bibinfo{person}{Jingyi Deng}, \bibinfo{person}{Jinlan Fu}, \bibinfo{person}{Kexin Huang}, {et~al\mbox{.}}} \bibinfo{year}{2024}\natexlab{}.
\newblock \showarticletitle{From gpt-4 to gemini and beyond: Assessing the landscape of mllms on generalizability, trustworthiness and causality through four modalities}.
\newblock \bibinfo{journal}{\emph{arXiv preprint arXiv:2401.15071}} (\bibinfo{year}{2024}).
\newblock


\bibitem[Maaz et~al\mbox{.}(2024)]%
        {maaz2023video}
\bibfield{author}{\bibinfo{person}{Muhammad Maaz}, \bibinfo{person}{Hanoona Rasheed}, \bibinfo{person}{Salman Khan}, {and} \bibinfo{person}{Fahad Khan}.} \bibinfo{year}{2024}\natexlab{}.
\newblock \showarticletitle{Video-ChatGPT: Towards Detailed Video Understanding via Large Vision and Language Models}. In \bibinfo{booktitle}{\emph{Proceedings of the 62nd Annual Meeting of the Association for Computational Linguistics (Volume 1: Long Papers)}}.
\newblock


\bibitem[Nihei et~al\mbox{.}(2024)]%
        {nihei2024chatbots}
\bibfield{author}{\bibinfo{person}{Misato Nihei}, \bibinfo{person}{Taiga Nohara}, \bibinfo{person}{Ikuko Sugawara}, {and} \bibinfo{person}{Takazumi Ono}.} \bibinfo{year}{2024}\natexlab{}.
\newblock \showarticletitle{Chatbots as Tools in Parent--Child Relationships}. In \bibinfo{booktitle}{\emph{International Conference on Human-Computer Interaction}}. Springer, \bibinfo{pages}{228--241}.
\newblock


\bibitem[O'Callaghan et~al\mbox{.}(2005)]%
        {o2005barriers}
\bibfield{author}{\bibinfo{person}{Anna~M O'Callaghan}, \bibinfo{person}{Lindy McAllister}, {and} \bibinfo{person}{Linda Wilson}.} \bibinfo{year}{2005}\natexlab{}.
\newblock \showarticletitle{Barriers to accessing rural paediatric speech pathology services: Health care consumers’ perspectives}.
\newblock \bibinfo{journal}{\emph{Australian Journal of Rural Health}} \bibinfo{volume}{13}, \bibinfo{number}{3} (\bibinfo{year}{2005}), \bibinfo{pages}{162--171}.
\newblock


\bibitem[Pepper and Weitzman(2004)]%
        {pepper2004talk}
\bibfield{author}{\bibinfo{person}{Jan Pepper} {and} \bibinfo{person}{Elaine Weitzman}.} \bibinfo{year}{2004}\natexlab{}.
\newblock \bibinfo{booktitle}{\emph{It Takes Two to Talk: A Practical Guide for Parents of Children with Language Delays}}.
\newblock \bibinfo{publisher}{Hanen Centre}.
\newblock
\newblock
\shownote{Shows parents how to help their child communicate and learn language during everyday activities.}.


\bibitem[Song et~al\mbox{.}(2016)]%
        {song2016talklime}
\bibfield{author}{\bibinfo{person}{Seokwoo Song}, \bibinfo{person}{Seungho Kim}, \bibinfo{person}{John Kim}, \bibinfo{person}{Wonjeong Park}, {and} \bibinfo{person}{Dongsun Yim}.} \bibinfo{year}{2016}\natexlab{}.
\newblock \showarticletitle{TalkLIME: mobile system intervention to improve parent-child interaction for children with language delay}. In \bibinfo{booktitle}{\emph{Proceedings of the 2016 ACM International Joint Conference on Pervasive and Ubiquitous Computing}}. \bibinfo{pages}{304--315}.
\newblock


\bibitem[Sussman(1999)]%
        {sussman1999words}
\bibfield{author}{\bibinfo{person}{Fern Sussman}.} \bibinfo{year}{1999}\natexlab{}.
\newblock \bibinfo{booktitle}{\emph{More Than Words: A Guide to Helping Parents Promote Communication and Social Skills in Children with Autism Spectrum Disorder}}.
\newblock \bibinfo{publisher}{Hanen Centre}.
\newblock
\newblock
\shownote{Step by step guide for parents of preschool children with autism spectrum disorder and other social communication difficulties.}.


\bibitem[Tanneberg et~al\mbox{.}(2024)]%
        {tanneberg2024attentivesupport}
\bibfield{author}{\bibinfo{person}{Daniel Tanneberg}, \bibinfo{person}{Felix Ocker}, \bibinfo{person}{Stephan Hasler}, \bibinfo{person}{Joerg Deigmoeller}, \bibinfo{person}{Anna Belardinelli}, \bibinfo{person}{Chao Wang}, \bibinfo{person}{Heiko Wersing}, \bibinfo{person}{Bernhard Sendhoff}, {and} \bibinfo{person}{Michael Gienger}.} \bibinfo{year}{2024}\natexlab{}.
\newblock \showarticletitle{To Help or Not to Help: LLM-based Attentive Support for Human-Robot Group Interactions}. In \bibinfo{booktitle}{\emph{2024 IEEE/RSJ International Conference on Intelligent Robots and Systems (IROS)}}. \bibinfo{pages}{9130--9137}.
\newblock
\urldef\tempurl%
\url{https://doi.org/10.1109/IROS58592.2024.10801517}
\showDOI{\tempurl}


\bibitem[{The Interdisciplinary Council on Development and Learning}(2025)]%
        {dir_floortime}
\bibfield{author}{\bibinfo{person}{{The Interdisciplinary Council on Development and Learning}}.} \bibinfo{year}{2025}\natexlab{}.
\newblock \bibinfo{booktitle}{\emph{DIR Floortime}}.
\newblock
\urldef\tempurl%
\url{https://www.icdl.com/dir/floortime}
\showURL{%
\tempurl}


\bibitem[Yu et~al\mbox{.}(2024)]%
        {cyu2025identifying}
\bibfield{author}{\bibinfo{person}{Qian Yu}, \bibinfo{person}{Zong Ke}, \bibinfo{person}{Guofu Xiong}, \bibinfo{person}{Yu Cheng}, {and} \bibinfo{person}{Xiaojun Guo}.} \bibinfo{year}{2024}\natexlab{}.
\newblock \showarticletitle{Identifying money laundering risks in digital asset transactions based on ai algorithms}.
\newblock \bibinfo{journal}{\emph{Available at SSRN 5129145}} (\bibinfo{year}{2024}).
\newblock


\end{thebibliography}

\appendix

\end{document}
\end{document}